\documentclass[eqsecnum,showkeys,showpacs,nofootinbib,aps,epsfig]{revtex4}

\usepackage{graphicx}
\def\bea{\begin{eqnarray}}
\def\eea{\end{eqnarray}}

\newcommand{\nn}{\nonumber}

\def\beq{\begin{equation}}
\def\eeq{\end{equation}}

\newbox\pippobox

\begin{document}
\title{Coupling of Brans-Dicke scalar field with Horava-Lifshitz gravity}
\author{Joohan Lee}
\email{joohan@kerr.uos.ac.kr}
\affiliation{Department of Physics,
University of Seoul, Seoul 130-743 Korea}
\author{Tae Hoon Lee}\email{thlee@ssu.ac.kr}
\affiliation{Department of Physics and Institute of Natural
Sciences,\\ Soongsil University, Seoul 156-743 Korea}
\author{Phillial Oh}
\email{ploh@newton.skku.ac.kr}\affiliation{Department of Physics
and Institute of Basic Science, Sungkyunkwan University, Suwon
440-746 Korea}
\date{\today}
\begin{abstract}
We look for a Brans-Dicke type generalization of Horava-Lifshitz
gravity. It is shown that such a generalization is possible within
the detailed balance condition. Classically, the resulting theory
reduces in the low energy limit to the usual Brans-Dicke theory with
a negative cosmological constant for certain values of parameters.
We then consider homogeneous, isotropic cosmology and study the
effects of the new terms appearing in the model.
\end{abstract}
\keywords{Horava-Lifshitz gravity; Brans-Dicke theory}
\maketitle

\section{Introduction}
Recently, a new theory of gravity has been proposed by
Horava\cite{ph1, ph2, ph3}. This theory, being based on anisotropic
scaling of space and time, breaks the spacetime symmetry. It has a
much better UV behavior than the theories with the spacetime
diffeomorphism symmetry, but expected to reduce to Einstein's
gravity in the infrared limit recovering the spacetime
diffeomorphism symmetry\footnote{This is still an open question. For
problems raised and some of the possible ways of their resolution
see, for instance, \cite{Blas, Bellorin}.}. Physical constants such
as the speed of light, Newton's constant, and cosmological constant
all emerge from the relevant deformation of the non-relativistic
theory at short distance. These interesting features as well as
other related findings have received a great deal of
attention\cite{gc, HLG}.

On the other hand, there are many alternative theories and
extensions of the Einstein theory. In particular, various gravity
models with scalar fields have been considered in the context of
cosmology to explain the behaviors of the universe in the early
stage as well as in the late stage\cite{QE, KE}. Therefore, it would
be interesting to consider similar extensions in the context of
Horava's theory. In this regard, of particular interest is the one
with a non-minimally coupled scalar field\footnote{Minimally coupled
scalar source had already been investigated\cite{ph1, MCS}}, typical
examples being the Brans-Dicke field\cite{BD} and the dilaton
field\cite{DG}.

In this paper, we take the Brans-Dicke field as a concrete example
of the non-minimally coupled scalar field and consider its inclusion
into the framework of Horava's gravity. Originally, Horava
introduced the concept of the detailed balance condition as a way of
reducing the choice of the potential, motivated by analogous methods
used in quantum critical systems. Recently, many serious problems
were reported\cite{CBS} associated with strictly imposing this
condition. However, some of the problems can be alleviated by softly
breaking the condition. Although the fate of the detailed balance
condition remains to be seen, it will be interesting to see if the
detailed balance condition can be maintained when we try to
non-minimally couple the scalar field to the Horava-Lifshitz
gravity.

It turns out that such an extension is possible and it reduces to
the four-dimensional Brans-Dicke theory with negative cosmological
constant when only the lowest order derivative terms are kept and
parameters of the theory are chosen to satisfy certain conditions.

We then study cosmological implication of the theory assuming
homogeneity and isotropy, and including the curvature-squared terms.
Because of the symmetries these higher order terms become a single
term proportional to $a^{-4}$ which can be regarded as the radiation
with negative energy. However, it is not strictly so because the
normal matter would couple with the inverse of the Brans-Dicke
scalar field. We concentrate only on their effects on cosmology. We
find several interesting features. We discuss them in Sec. 3.

\section{Construction of the model}
Let us consider the four-dimensional Brans-Dicke theory\cite{BD},
where the action is given by \beq S=\int d^4x\sqrt{-g}\left(\phi
R-\omega\phi^{-1}g^{\mu\nu}\partial_\mu\phi\partial_\nu\phi\right).
\eeq Decomposition of this action into $(3+1)$ form, including the
speed of light, $c$, yields (See Ref. \cite{bk}, for instance.) \bea
\sqrt{-g}\phi R &\simeq& N\sqrt{q}\phi\left({\cal
R}+c^{-2}(K_{ab}K^{ab}-K^2)\right)
-2N\sqrt{q}c^{-2}K\pi-2N\sqrt{q}D^2\phi,\\
-\sqrt{-g}\omega\phi^{-1}
g^{\mu\nu}\partial_\mu\phi\partial_\nu\phi&=&N\sqrt{q}\omega\phi^{-1}c^{-2}\pi^2
-N\sqrt{q}\omega\phi^{-1}D^a\phi D_a\phi, \eea where the four metric
$g$ is decomposed into the lapse function $N$, the shift vector
$N^a$ and the three metric $q_{ab}$, and the corresponding
three-dimensional covariant derivative and its scalar curvature are
denoted respectively by $D_a$, ${\cal R}$. The Brans-Dicke parameter
is assumed positive, $\omega>0$. In the first equation irrelevant
total divergence terms were dropped out. The time derivatives of the
three-metric and the scalar field are encoded in the following
quantities; \bea K_{ab}&\equiv&{1\over
2N}({\dot g}_{ab}-D_aN_b-D_bN_a),\\
\pi&\equiv&{1\over N}({\dot \phi}-N^a\partial_a\phi).\eea Using the
above result the Brans-Dicke action can be split into the two parts
$S_{BD}=S_{BD}^K+S_{BD}^V$, where the kinetic and potential parts
can be written after re-scaling of the scalar field $\phi$ and the
corresponding field $\pi$ as
\bea S_{BD}^K&=&\int dt\,d^3x
N\sqrt{q}\left(
\phi(K_{ab}K^{ab}-K^2)-2K\pi+\omega\phi^{-1}\pi^2\right),\\
S_{BD}^V&=&c^2\int dt\,d^3x N\sqrt{q}\left(\phi {\cal
R}-2D^2\phi-\omega\phi^{-1}D^a\phi D_a\phi\right).\eea Note the
factor of $c^2$ in front of the potential term. For the later
purpose regarding the detailed balance condition it is important to
express the kinetic part in the following matrix form; \beq
S_{BD}^K=\int dt\,d^3x N\sqrt{q} \left(
  \begin{array}{cc}
    K_{ab} & \pi \\
  \end{array}
\right) \left(\begin{array}{cc}
      \phi G^{abcd} & -q^{ab} \\
      -q^{cd} & \omega\phi^{-1} \\
       \end{array}\right)
       \left(\begin{array}{c}
      K_{cd} \\
      \pi \\
      \end{array}\right),
 \eeq where \beq G^{abcd}=
\frac{1}{2}\left(q^{ac}q^{bd}+q^{ad}q^{bc}\right)-q^{ab}q^{cd}.\label{sm}
\eeq The matrix in the middle of the kinetic part of the action can
be regarded as the supermetric on the space of $(q_{ab},\phi)$,
naturally extending the DeWitt metric on the space of three-metrics.

We intend to construct a Brans-Dicke type extension of
Horava-Lifshitz gravity with the detailed balance condition. So, we
choose the action of the form, $S_{HLBD}=S_{HLBD}^K+S_{HLBD}^V$,
where the kinetic part is \beq S_{HLBD}^K=\int dt d^3x N\sqrt{q}
\left(
  \begin{array}{cc}
    K_{ab} & \pi \\
  \end{array}
\right) \left(\begin{array}{cc}
      \phi G^{abcd}(\lambda) & -q^{ab} \\
      -q^{cd} & \omega\phi^{-1} \\
       \end{array}\right)
       \left(\begin{array}{c}
      K_{cd} \\
      \pi \\
      \end{array}\right)\eeq
and the potential part is of the form \beq S_{HLBD}^V=-\int dt d^3x
N\sqrt{q} \left(
  \begin{array}{cc}
    \frac{\delta W}{\delta q_{ab}} & \frac{1}{2}\frac{\delta W}{\delta\phi} \\
  \end{array}
\right) {\left(\begin{array}{cc}
      \phi G^{abcd}(\lambda) & -q^{ab} \\
      -q^{cd} & \omega\phi^{-1} \\
       \end{array}\right)}^{-1}\left(\begin{array}{c}
      \frac{\delta W}{\delta q_{cd}} \\
      \frac{1}{2}\frac{\delta W}{\delta\phi} \\
      \end{array}\right)\label{qqq}\eeq
for some suitable choice of function $W(q,\phi)$. The supermetric
$G^{abcd}(\lambda)$ was slightly deformed compared to the Eq.
(\ref{sm}) to include the parameter $\lambda$ as usual, \beq
G^{abcd}(\lambda)\equiv
\frac{1}{2}\left(q^{ac}q^{bd}+q^{ad}q^{bc}\right)-\lambda
q^{ab}q^{cd}.\eeq The factor of two was inserted in front of the
variation of $W$ with respect to $\phi$ to compensate for different
normalization in time derivatives in Eqs.(4) and (5). It is a
straightforward matter to calculate the inverse supermetric. It
comes out to be of form \beq \left(\begin{array}{cc}
      \phi^{-1} {\cal G}_{abcd} & -Aq_{ab} \\
      -Aq_{cd} & B\phi \\
       \end{array}\right),\eeq
where \beq {\cal G}_{abcd}=
\frac{1}{2}\left(q_{ac}q_{bd}+q_{ad}q_{bc}\right)-\bar{\lambda}q_{ab}q_{cd},
\eeq with \beq A=\frac{1}{\omega(3\lambda-1)+3},~~
B=\frac{3\lambda-1}{\omega(3\lambda-1)+3},~~
\bar{\lambda}=\frac{1+\omega\lambda}{\omega(3\lambda-1)+3}.\eeq Note
that this inverse supermetric is well-defined even for $\lambda=1/3$
contrary to the pure gravity case and becomes singular instead when
$\lambda=(\omega-3)/3\omega$, for instance when $\lambda=1$ and
$\omega=-3/2$ corresponding to the conformal scalar case (We assume
$\omega>0$ in this work.). If we take the limit of
$\omega\rightarrow\infty$, $A$ and $B$ vanish and
$\bar{\lambda}=\lambda/(3\lambda-1)$, reproducing the pure gravity
case.

We choose \beq W=c_1\int d^3x \sqrt{q}\phi ({\cal
R}-2\Lambda_b)-c_2\int d^3x\sqrt{q}\omega\phi^{-1}D^a\phi
D_a\phi.\eeq In general all possible marginal and relevant terms can
be included. The above choice of $W$ corresponds to keeping only
terms important in the infrared limit. Then, after a straightforward
calculation Eq. (\ref{qqq}) can be written as  \bea
S_{HLBD}^V&=&\int dt\,d^3xN\sqrt{q}\left\{\alpha
\phi+\beta(\phi{\cal R}-\frac{c_2}{c_1}\omega\phi^{-1}D^a\phi
D_a\phi)
+\gamma(-2D^2\phi)\right\}\nn\\
&-&\int dt\,d^3xN\sqrt{q}\left(Q^{ab}\phi^{-1}{\cal
G}_{abcd}Q^{cd}-2AQ^{ab}q_{ab}Q+B\phi Q^2\right),\label{quad}\eea
where \bea
\alpha&=&(c_1\Lambda_b)^2\frac{3\omega+7-3\lambda}{\omega(3\lambda-1)+3}\\
\beta&=&-(c_1)^2\Lambda_b\frac{\omega+5-3\lambda}{\omega(3\lambda-1)+3}\\
\gamma&=&-(c_1)^2\Lambda_b\frac{2(\omega+1)
-\frac{c_2}{c_1}\omega(4-3\lambda)}{\omega(3\lambda-1)+3},\eea and
\bea Q^{ab}&\equiv& c_1\left(-\phi
({\cal R}^{ab}-\frac{1}{2}{\cal R}q^{ab})+D^aD^b\phi-q^{ab}D^2\phi\right),\\
Q&\equiv& c_1\frac{{\cal
R}}{2}-c_2\left(-\omega\phi^{-1}D^2\phi+\frac{\omega}{2}\phi^{-2}D^a\phi
D_a\phi\right).\eea The second line of Eq. (\ref{quad}) has
quadratic terms only.

When $c_1=c_2$ and $\lambda=1$, the theory recovers four-dimensional
diffeomorphism symmetry, as one can see from the fact that in the
infrared limit the potential part of the action becomes \beq
S_{BDHL}^V\vert_{IR}=
-(c_1)^2\Lambda_b\frac{\omega+2}{2\omega+3}\int dt\,d^3x
N\sqrt{q}\left(\phi ({\cal
R}-2\Lambda)-2D^2\phi-\omega\phi^{-1}D^a\phi D_a\phi\right),\eeq
where \beq \Lambda=\frac{3\omega+4}{2(\omega+2)}\Lambda_b.\eeq This
expression coincides with that of the Brans-Dicke theory except that
the cosmological constant term is present. Comparison with the
kinetic part yields the speed of light \beq
c^2=-(c_1)^2\Lambda_b\frac{\omega+2}{2\omega+3}.\eeq As in the case
of the Horava gravity the constant $\Lambda_b$ must be negative,
consequently allowing only negative cosmological constant $\Lambda$.
The Newton constant is related to the scalar field $\phi$ as
follows,\beq G_N=\frac{c^2}{16\pi\phi}.\eeq

\section{Cosmological solutions}
Now, we consider the homogeneous, isotropic cosmology. We restrict
ourself to the case of $\lambda=1$, $c_1=c_2$, and set the speed of
light to unity, i.e., $c=1$. We choose vanishing shift vector
$N^a=0$, and the three-metric to be the usual maximally symmetric
ones with curvature constant $k=-1,0,+1,$ \beq ds^2=
a^2(t)\left(\frac{dr^2}{1-kr^2}+r^2(d\theta^2+\sin^2\theta
d\phi^2)\right).\eeq In this case the higher derivative terms become
greatly simplified due to homogeneity and isotropy, \beq
Q^{ab}=kc_1\frac{\phi}{a^2} q^{ab},~~~~ Q=6kc_1\frac{1}{a^2}.\eeq
The field equations become \bea 3
H^2+3H\frac{\dot{\phi}}{\phi}-\frac{1}{2}
\omega(\frac{\dot{\phi}}{\phi})^2&=&\frac{1}{2}\phi^{-1}\rho_m-\frac{3k}{a^2}+\Lambda
-\frac{1}{2}(\frac{B^2}{a^4}),\nn\\
-2\dot{H}-3H^2-\frac{\ddot{\phi}}{\phi}-2H\frac{\dot{\phi}}{\phi}
-\frac{\omega}{2}(\frac{\dot{\phi}}{\phi}))^2&=&\frac{1}{2}\phi^{-1}p_m+\frac{k}{a^2}-\Lambda-
\frac{1}{6}(\frac{B^2}{a^4}),\nn\\
(2\omega+3)\left(\frac{\ddot{\phi}}{\phi}
+3H\frac{\dot{\phi}}{\phi}\right)&=&\frac{1}{2}\phi^{-1}(\rho_m-3p_m)+2\Lambda
+\frac{B^2}{a^4},\label{fried}\eea together with the usual form of
the continuity equation for the matter density $\rho_m$ for
consistency, where $H\equiv(\dot{a}/{a})$ is the Hubble constant and
\beq
B^2=\frac{3\omega}{2\omega+3}(kc_1)^2=\frac{3\omega(3\omega+4)}{2(\omega+2)}
\frac{k^2}{(-\Lambda)}.\eeq The first equation in Eq. (\ref{fried})
is the Friedmann equation of the Brans-Dicke theory with a negative
cosmological term and the dark radiation term included. Again, we
emphasize that they do not have $\phi^{-1}$ coupling in contrast to
the normal matter. In the absence of those two terms the equations
simply become those of the usual Brans-Dicke theory\cite{HKim}.
Therefore, we restrict our attention to the new effects resulting
from those two terms.

For simplicity, assume that the matter is absent, i.e.,
$\rho_m=p_m=0$. First, consider the case where the dark radiation
like term dominates, so Eq. (\ref{fried}) reduces to \bea 3
H^2+3H\frac{\dot{\phi}}{\phi}-\frac{1}{2}
\omega(\frac{\dot{\phi}}{\phi})^2&=&-
\frac{1}{2}(\frac{B^2}{a^4}),\nn\\-2\dot{H}-3H^2-\frac{\ddot{\phi}}{\phi}-2H\frac{\dot{\phi}}{\phi}
-\frac{\omega}{2}(\frac{\dot{\phi}}{\phi}))^2&=&
-\frac{1}{6}(\frac{B^2}{a^4}),\nn\\
(2\omega+3)\left(\frac{\ddot{\phi}}{\phi}
+3H\frac{\dot{\phi}}{\phi}\right)&=& \frac{B^2}{a^4}.\eea To further
simplify these equation we set \beq X\equiv H+Y,~~~~
Y\equiv\frac{1}{2}\left(\frac{\dot{\phi}}{\phi}\right),\eeq in terms
of which, they can be written as
\bea &3X^2&-AY^2=-\frac{B^2}{2a^4},\nn\\
&\dot{X}&=-\frac{3}{2}X^2+XY-\frac{A}{2}Y^2+\frac{1}{6}(\frac{B^2}{2a^4}),\nn\\
&A\dot{Y}&=-3AXY+AY^2+\frac{B^2}{2a^4},\eea where
$A\equiv2\omega+3$. Setting \bea
\sqrt{3}X&=&\sqrt{\frac{B^2}{2a^4}}~~{\rm{sinh}}\theta,\nn\\
\sqrt{A}Y&=&\sqrt{\frac{B^2}{2a^4}}~~{\rm{cosh}}\theta,\eea Eq.
(\ref{fried}) yields \beq
\dot{\theta}=-\sqrt{\frac{B^2}{2a^4}}\left(\frac{1}{\sqrt{3}}{\rm{cosh}}~\theta
+\frac{1}{\sqrt{A}}{\rm{sinh}}~\theta\right).\eeq Note that
$\dot{\theta}$ is always negative for $\omega>0$, which means that
$\theta$ goes from negative infinity to positive infinity as time
flows. Combine this result with \beq
\frac{\dot{a}}{a}=X-Y=\sqrt{\frac{B^2}{2a^4}}\left(-\frac{1}{\sqrt{A}}{\rm{cosh}}!\theta
+\frac{1}{\sqrt{3}}{\rm{sinh}}~\theta\right)\eeq to get \beq
\frac{d{\rm log}a}{d\theta}=-\left(\frac{
\frac{1}{\sqrt{3}}{\rm{sinh}}~\theta-\frac{1}{\sqrt{A}}{\rm{cosh}}~\theta}{\frac{1}{\sqrt{3}}{\rm{cosh}}~\theta
+\frac{1}{\sqrt{A}}{\rm{sinh}}~\theta}\right).\eeq Although this
equation can be integrated, the resulting expressions can be quite
complicated, asymptotic behavior of $a(t)$ at early and late times
can be easily determined. A straightforward analysis shows that the
scale factor vanishes at initial time and finite later time. The
universe it describes expansion from a singularity and within a
finite time it collapses.

When the cosmological term dominates, Eq. (\ref{fried}) becomes
\bea &3X^2&-AY^2=\Lambda,\nn\\
&\dot{X}&=-\frac{3}{2}X^2+XY-\frac{A}{2}Y^2+\frac{\Lambda}{2}\nn\\
&A\dot{Y}&=-3AXY+AY^2+\Lambda.\eea With \bea
\sqrt{3}X&=&\sqrt{-\Lambda}~~{\rm{sinh}}~\theta,\nn\\
\sqrt{A}Y&=&\sqrt{-\Lambda}~~{\rm{cosh}}~\theta,\eea we find a
slightly different equation for $\theta$, \beq
\dot{\theta}=-\sqrt{-\Lambda}\left(\sqrt{3}{\rm{cosh}}~\theta
-\frac{1}{\sqrt{A}}{\rm{sinh}}~\theta\right).\eeq General behavior
of the solution is the same as the previous case.

\section{Conclusion and discussion}
To summarize, we constructed a Brans-Dicke type extension of the
Horava-Lifshitz gravity maintaining the detailed balance condition.
Although strict imposition of the detailed balance condition is
known to have many problems, one can either break the condition or
simply treat the resulting terms as an important contribution to the
potential. We have not discussed the issue of projectability in this
work. At this level, our model can be incorporated into any version.

Furthermore, We investigated its low energy limit and shown that the
resulting theory is the Brans-Dicke theory with a negative
cosmological constant. In Brans-Dicke theory one can incorporate
cosmological constant term in two ways. One is treating it as a
vacuum expectation value from the matter sector and the other is
what we have done in this work.

We studied the curvature contribution up to quadratic order in the
context of homogeneous and isotropic cosmology. The resulting theory
is the Brans-Dicke theory with a negative cosmological constant and
a radiation-like term with a negative energy density. They are
somewhat different from usual matter in that they do not have
$\phi^{-1}$ coupling. We analyzed their effects and showed that the
resulting solution has the general behavior of big rip. This is in
contrast with the pure gravity case of Horava.

Although we focused on the Brans-Dicke theory in this paper the
analysis can be generalized to other non-minimally coupled scalar
field gravity theory. It would be interesting to further investigate
cosmological aspects of the resulting theories.

\acknowledgments THL was supported by the Soongsil University
Research Fund. PO was supported by the National Research Foundation
of Korea(NRF) grant funded by the Korea government(MEST) through the
Center for Quantum Spacetime(CQUeST) of Sogang University with grant
number 2005-0049409.

\end{document}